\documentclass{iau} 
\usepackage{graphicx}
\usepackage{upgreek}
\usepackage{eurosym}
\usepackage{natbib}

\title[Pulsars with the SKA] 
{Pulsar Science with the SKA}

\author[Evan F. Keane]   
{Evan F. Keane$^1$}

\affiliation{$^1$SKA Organisation, Jodrell Bank Observatory,
  Macclesfield, Cheshire, SK11 9DL, UK \\ email: {\tt
  e.keane@skatelescope.org}}

\pubyear{2017}
\volume{337}  
\jname{Pulsar Astrophysics -- The Next 50 Years}
\editors{P. Weltevrede, B.B.P. Perera, L. Levin Preston \& S. Sanidas, eds.}

\begin{document}

\maketitle

\begin{abstract}
  The Square Kilometre Array (SKA) will be sensitive enough to
  discover all of the pulsars in the Milky Way that are beamed towards
  Earth. Already in the initial deployment, SKA Phase 1, it will make
  significant advances in pulsar science. In these proceedings I
  briefly overview what the SKA is, and describe its pulsar search and
  timing capabilities.
\end{abstract}

\firstsection 
\section{What is the SKA?}
What \textit{is} the SKA? This oft-posed question usually results in a
wide range of answers. According to publishable responses made by the
delegates of IAU 337, the SKA is:``crazy'', ``crazy, but in a good
way'', ``the most ambitious science project there has ever been'',
``the future of pulsar astronomy'', ``imminent'', ``bloody
difficult'', ``actually coming together''. The SKA is a number of
things, which we now expand upon.

\subsection{Two Telescopes}
\begin{figure*}
  \begin{center}
    \includegraphics[scale=0.3, trim = 0mm 0mm 0mm 0mm, clip]{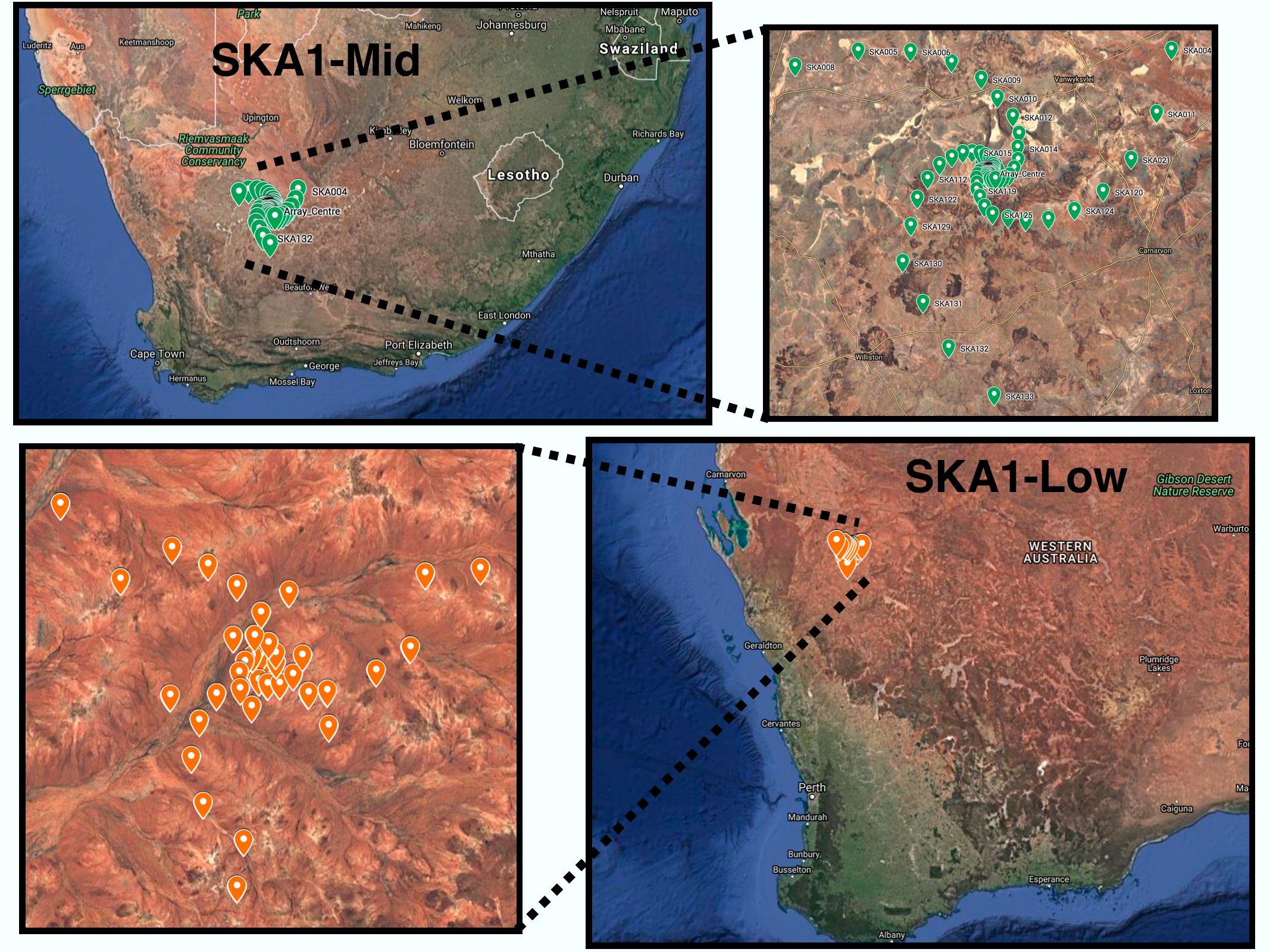}
    \caption{\scriptsize{The locations of the initial deployment
        SKA1-Low and SKA1-Mid arrays overlayed on a Google Maps
        display (Map data: \textcopyright2017 AfriGIS (Pty) Ltd,
        Google Imagery \textcopyright2017 NASA, TerraMetrics). The top
        left panel shows the location of the SKA1-Mid array in South
        Africa. The top right inset zooms in on the individual
        SKA1-Mid dishes; the three spiral arms of dishes are evident
        but the elements in the core are unresolved at this scale due
        to the density of dishes in the inner region. The bottom right
        panel shows the location of the SKA1-Low array in Western
        Australia. The bottom left inset zooms in on the individual
        SKA1-Low stations; the three-arm spiral is perturbed due to
        constraints arising from the terrain and land ownership. As
        for SKA1-Mid the core stations are unresolved at this zoom
        level due to the density in the core. Pulsar applications will
        typically use one or more sub-arrays consisting of elements in
        the high-sensitivity core regions. This map is accessible via
        \texttt{https://github.com/evanocathain/SKA}.}}\label{fig:sites}
  \end{center}
\end{figure*}

\textit{The SKA is 2 telescopes}: The SKA observatory consists of two
array telescopes, each located at a different site and each covering a
different sky frequency range. The initial deployment of SKA, termed
Phase 1 or simply SKA1, consists of a `low' frequency array in Western
Australia and a `mid' frequency array in South Africa (see
Figure~\ref{fig:sites}, Table~\ref{tab:bands}).
A number of SKA precursor instruments are currently
operating at these sites: the Murchison Widefield Array (MWA) and
Australian SKA Pathfinder (ASKAP) in Western Australia; the Hydrogen
Epoch of Reionization Array (HERA) and MeerKAT in South
Africa. MeerKAT will be fully integrated into the SKA1-Mid array;
SKA1-Mid thus consists of 133$\times$15-m `SKA' dishes and
64$\times$13.5-m MeerKAT dishes.

To justify its name the full SKA deployment, often termed SKA2, will
need to have a physical collecting area of $1\;\mathrm{km}^2$ for each
telescope which is equivalent to that of a $\sim 1130$-m single dish
fully illuminated. In terms of how close the initial deployment is to
this we note that SKA1-Mid has a physical collecting area of
$0.033$~km$^2$, equivalent to a single dish diameter of 200
metres. 
SKA1-Low is an aperture array consisting of $512$ stations of $256$
log-periodic dipole antennas, so that the collecting area is frequency
dependent. It falls as the square of the sky frequency above the
`dense-sparse' transition frequency which is at approximately
$100$~MHz for the SKA1-Low design; thus at zenith SKA1-Low has an
effective collecting area of $\sim 0.7$~km$^2$ at 100 MHz, but only
$\sim 0.2$~km$^2$ and $\sim 0.1$~km$^2$ at 200 and 300 MHz
respectively.
Clearly the jump from SKA1 to SKA2 will need to be larger for Mid than
for Low; indeed the full deployment of SKA-Mid is planned to expand
into eight more countries in Southern Africa\footnote{The African
  partner countries are Botswana, Ghana, Kenya, Madagascar, Mauritius,
  Mozambique, Namibia and Zambia.}.

\begin{table}
  \centering
  \caption{\scriptsize{The frequency coverage of SKA1. The SKA1-Low
      antenna design has a 7:1 frequency ratio and involves a
      log-periodic dipole. For SKA1-Mid, Band 1 is a 3:1 system but
      the others are all octave bands. SKA1-Mid bands marked with an
      asterisk are, at present, not funded in the initial construction
      budget agreed by the member countries (674 M\euro, in units of
      April-2016 euros). Additionally there are wide-band single-pixel
      feeds with preliminary designs known as Band A (which
      encompasses the frequency range of Bands 3 and 4) and Band B
      (which encompasses the entirety of Band 5). The dish cryostat
      has empty slots for Bands 3, 4, A (instead of 3), B (instead of
      5a) and 6 should they be deployed in an upgrade; another
      possible upgrade is to swap Band 1 for a phased array feed.}}
  \label{tab:bands}
  \begin{tabular}[h]{lll}
    \hline
    SKA1-Low & SKA1-Mid \\
    \hline
    $50-350$~MHz & Band 1: $350-1050$~MHz \\
                 & Band 2: $0.95-1.76$~GHz \\
                 & Band 3*: $1.65-3.05$~GHz \\
                 & Band 4*: $2.80-5.18$~GHz \\
                 & Band 5a: $4.6-8.5$~GHz \\
                 & Band 5b: $8.3-15.4$~GHz \\
                 & Band 6*: $13.5-25.0$~GHz \\
    \hline
  \end{tabular}
  \vspace{-1cm}
\end{table}

The top level design of SKA1 is frozen~\citep{L1_reqs}. At the time of
writing (October 2017) we are in the latter stages of the detailed
design phase for SKA1, which itself began in 2013. At that time the
entire SKA1 system was divided into design elements (and thence
sub-elements). The critical design reviews of the elements, and thence
the system, are scheduled to occur during 2018. At this time the
project progresses to the construction phase which is expected to take
5 years. As it is already a large leap in capabilities in many senses
compared to other facilities the observatory will operate with the
Phase 1 deployment for several years before expanding to the full SKA.


\subsection{Science Driven}
\textit{The SKA is science driven}: For SKA1 there are 13 High
Priority Science Objectives (HPSOs, see \citealt{L0_reqs}); one is
pulsar search, another pulsar timing. The initial deployment is
designed to deliver on these in particular but will be capable of much
more. The full SKA science case is extremely broad and described in
the 2015 Science Book~\citep{SKA_Book}, a 2000-page 1200-author
135-chapter tome on planned science activities with the
SKA. 
Amongst the HPSOs pulsar science is unique in its use of both Mid and
Low. Fast Radio Burst (FRB) science is considered a Mid HPSO simply
because
FRBs have not yet been detected in the Low band; in practice however
FRB searching comes with the ability to perform pulsar searches and
will also occur on SKA1-Low.


In the first 5 years the HPSOs will be addressed. The observing time
needed amounts to $\sim 3.7$~y and $\sim 9.9$~y for Low and Mid
respectively~\citep{L0_reqs}. Even considering just basic
observational constraints (e.g. Epoch of Reionisation observations can
only meaningfully be performed at night and when the ionospheric
conditions permit, pulsar timing applications need a lot of Galactic
time, etc.), down time, and the fraction of time spent on non-HPSO
science, it is clear that the HPSO time far exceeds wall time if
everything is done sequentially.
Fortunately commensality is an integral part of the design and one can
(on both Low and Mid) perform imaging, pulsar search and pulsar timing
simultaneously. Furthermore SKA1-Low can form multiple station beams
in different directions on the sky, with each performing different
(and multiple) types of observation.
On both Low and Mid there is a single pulse search which will identify
pulsars via their single pulses (rather than by periodicity searches),
but also FRBs. This mode is quite low power (power availability being
a key constraint at both sites) so that in principle it could be run
commensally 24/7 meaning these single pulse searches would come `for
free' (at least in the sense of observing time). Recently the ability
to accommodate `custom experiments' has been added to the
design. Subject to various practical constraints at each site one
could then `plug in' additional hardware and/or run custom pipelines
to add further capabilities for simultaneous science applications.

\subsection{Sovereign}
\textit{The SKA is sovereign}: The 10 SKA member countries which fund
the project are: Australia, Canada, China, India, Italy, the
Netherlands, New Zealand, South Africa, Sweden and the United
Kingdom. At present the SKA is transitioning, becoming an
inter-governmental organisation (IGO). Examples of other IGOs in
operation are ESO, CERN, NATO and the UN. The details of the IGO
treaty are currently being finalised at government level in the member
countries and ultimately require ratificiation in the various
parliaments. This IGO process is happening in parallel to the final
stages of the technical design work, and should conclude before
construction will begin.

\subsection{A Community}
\textit{The SKA is a community}: There is a large community of
astronomers world wide working on SKA science. The SKA Science Working
Groups (SWGs), and Focus Groups comprise $\sim 900$ professional
astronomers from around the world; this community has shaped what the
SKA is and is who the SKA will serve. The Pulsar SWG is amongst the
most active and engaged of all groups and is open to any active
PhD-level pulsar astronomer who is interested in contributing. The SWG
is organised into two tiers, with a smaller core group led by two
chairs\footnote{At the time of writing the Pulsar SWG chairs are
  Andrea Possenti and Ingrid Stairs; the author is the relevant point
  of contact in the SKA science team.}.

\section{Pulsar Capabilities}
Pulsar-driven design requirements are to be found throughout the
entire SKA1 system. Most intensive `non-image processing' is
concentrated within the Central Signal Processor (CSP) which is the
element for the beam-formers and correlators, but also the pulsar
search and timing back-ends. CSP is located at the observatory sites,
and fed with data from the individual dishes/stations (the SKA1-Low
stations have an additional station-level beam-former). Within CSP,
beam-forming occurs and the resultant data products go to the pulsar
search and/or timing back-ends. After CSP, the data are transported to
the Science Data Processor (SDP). Correlated data for imaging
applications are sent directly from the CSP correlator to SDP with no
further processing at the observatory site; imaging occurs in SDP. For
pulsar purposes SDP assesses pulsar candidates identified by the
pulsar search backend, and performs pulsar timing from the folded
profiles sent to it from the timing backend. Another element of
interest for pulsar applications is Signal And Data Transport (SaDT)
which is responsible for the observatory clocks, the distribution of
time and frequency across the arrays and the networking of all
elements.

\subsection{Pulsar Search Capabilities}\label{sec:search}
Pulsar search capabilities are provided by the pulsar search backend
(PSS), a sub-element of CSP. It can perform real-time pulsar searches
for observations with durations up to 30 mins. These searches
dedisperse the data for dispersion measures (DM) from $0$ to
$3000$~$\mathrm{pc}\;\mathrm{cm}^{-3}$. A single pulse search (for
pulsars and FRBs) is performed across this range. An acceleration
search is also performed: for a 10-minute pointing, this involves
acceleration searches for binary pulsars in the range $|a|<
350$~$\mathrm{m\;s}^{-2}$ for at least $500$ DM trials. This happens
for each tied-array power beam provided by the beamformer --- 500 on
SKA1-Low, 1500 on SKA1-Mid.
The bandwidth of these beams is $100$ and $300$~MHz respectively, and
one can have less beams for proportionally more bandwidth.

For pulsar applications the entire SKA1 arrays will not be used. The
improvement in sensitivity with the addition of long-baseline elements
is outweighed by the large decrease in beam-size, i.e. a much larger
beam-former would be needed. 
PSS can operate in up to 16 sub-arrays each containing elements within
a 20-km diameter. The maximum size pulsar sub-array (one centred at
the array centre) then consists of 404 stations and 164 dishes (100
SKA + all 64 MeerKAT) on SKA1-Low and -Mid respectively, corresponding
to $\sim 79\%$ and $\sim 82\%$ of the full array effective collecting
areas.
Pulsar search will most likely use a sub-array consisting of
approximately
the inner 1~km, with the 20-km sub-array option being used for a
$(20)^2=400$ times improvement on the initial localisation, as done
with LOFAR~\citep{Stappers_2011}.
For the inner $1$~km this then involves 224 stations and 93 dishes (55
SKA + 38 MeerKAT) on SKA1-Low and -Mid respectively, corresponding to
$\sim 44\%$ and $\sim 46\%$ of the full array effective collecting
areas.

\begin{figure*}
  \begin{center}
    \includegraphics[scale=0.5, trim = 10mm 20mm 0mm 10mm, clip]{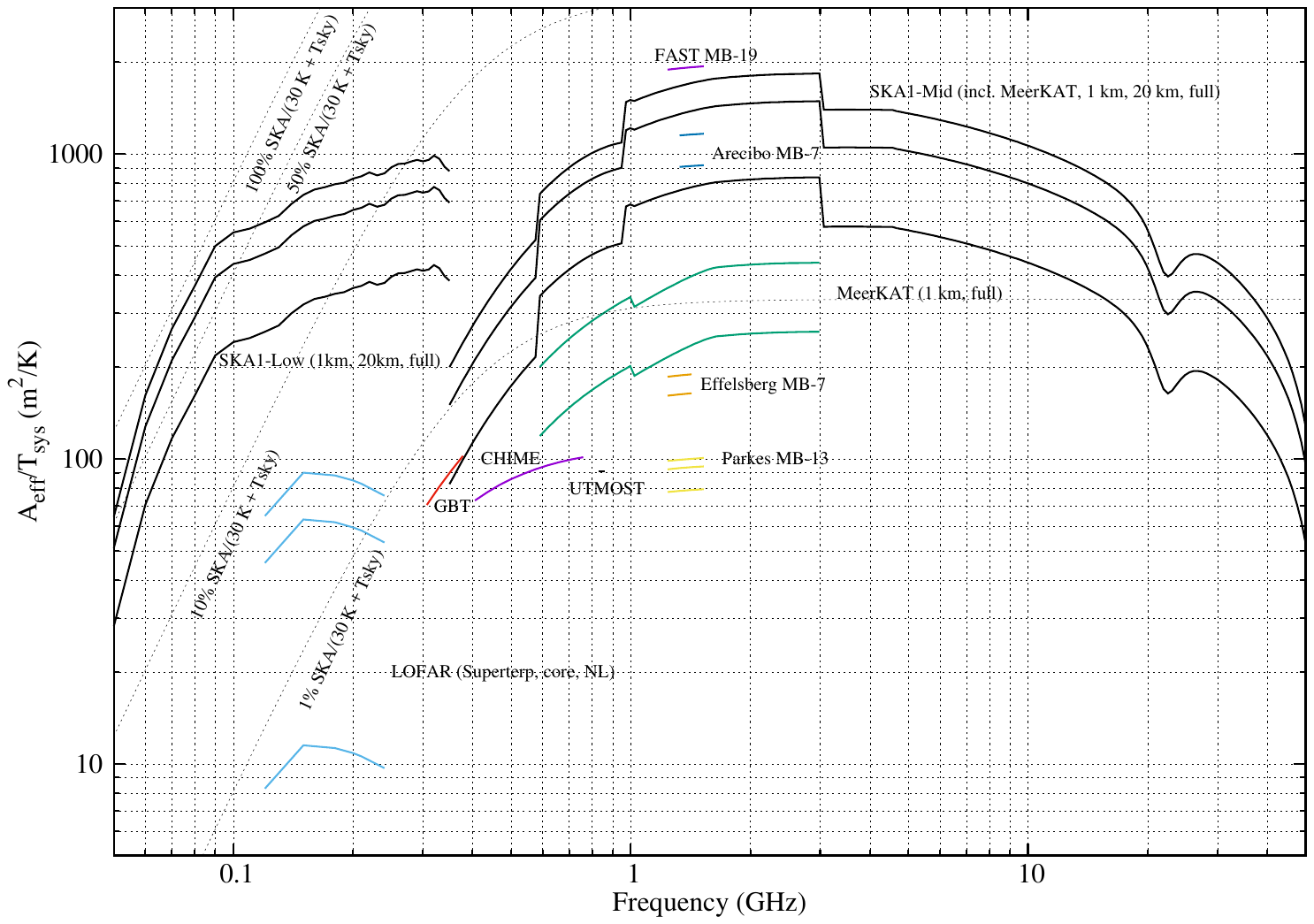}
    \includegraphics[scale=0.5, trim = 10mm 20mm 0mm 10mm, clip]{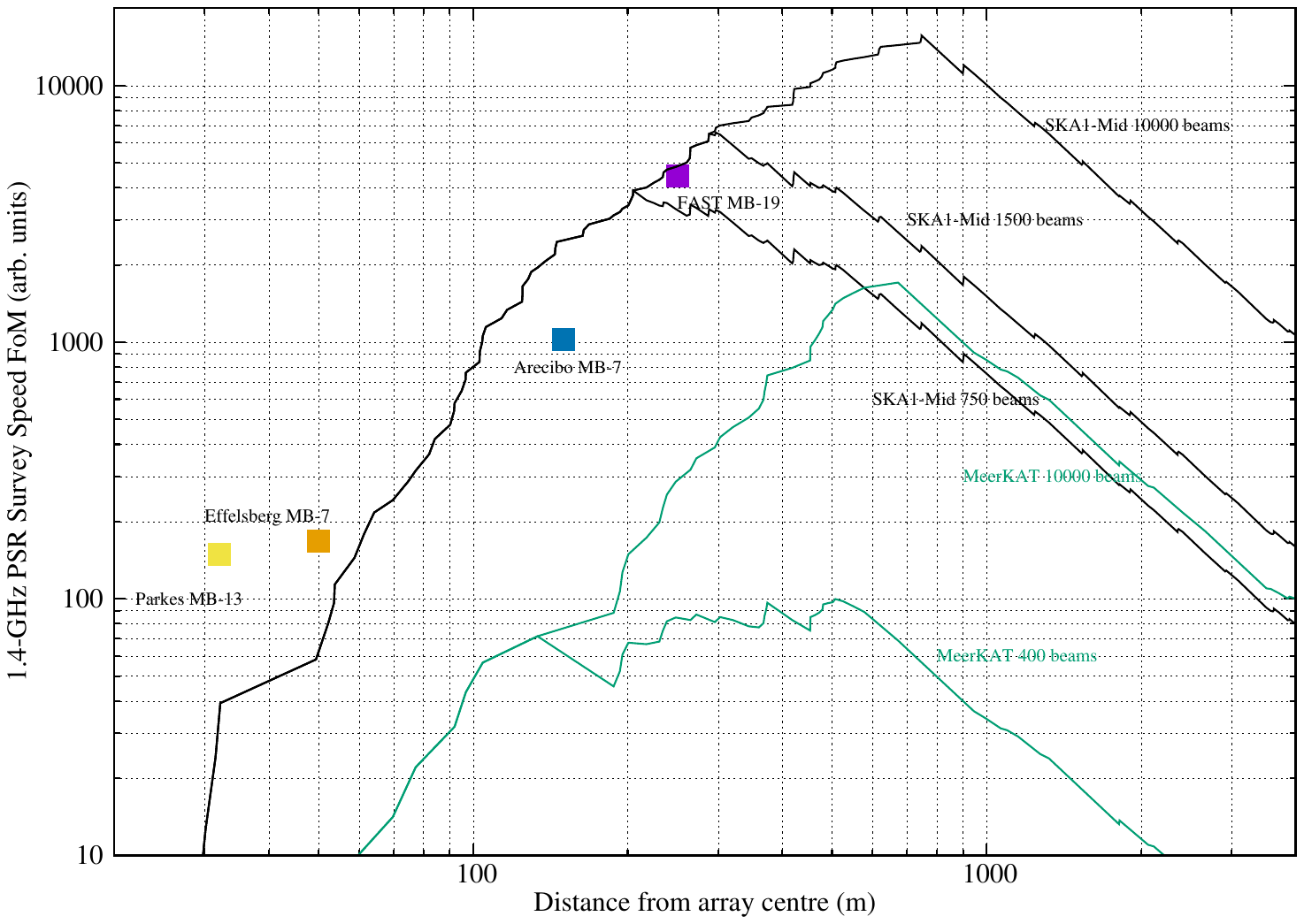}
    \caption{\scriptsize{Top: The gain of SKA1 as compared to some
        pulsar-search-relevant facilities. Three different curves for
        for the anticipated SKA1 performance are shown --- for
        sub-arrays of diameter 1~km, 20~km and the full array (see
        \citealt{performance} and references therein; this and other
        relevant SKA material can be found at
        \texttt{https://astronomers.skatelescope.org/documents}).
      The array configurations along with a basic sensitivity
      calculator can be found at
      \texttt{https://github.com/evanocathain/SKA} where one can
      obtain the sensitivity of SKA1 for a user-requested sub-array, sky location etc.
      Here the 50th percentile contributions of sky temperature and
      precipitable water vapour are used and the sensitivity at zenith
      is plotted --- for SKA1-Low the dependence is approximately
      $\propto\cos^{2}(z)$
      ~\citep{L1_reqs}. For FAST we show the anticipated performance
      of the 19-beam L-band receiver~\citep{li_pan}, for Arecibo the
      7-beam PALFA receiver~\citep{palfa}, for Effelsberg the 7-beam
      receiver used for the HTRU-N survey~\citep{htrun}, for Parkes
      the 13-beam receiver used for the HTRU-S and SUPERB
      surveys~\citep{htrus,superb}, for GBT the GBNCC
      specifications~\citep{gbncc}, for UTMOST the upgraded
      specifications~\citep{utmost}, for CHIME as estimated by
      \citet{rl17}, and for LOFAR as specified in
      \citet{vanhaarlem}. 
      The anticipated SKA1-Mid dish performance is shown to $50$~GHz,
      beyond the frequency range where bands are currently
      planned. Bottom: The equal-bandwidth pulsar survey speed figure
      of merit at $1.4$~GHz for SKA1-Mid as compared to a selection of
      other instruments. SKA1-Mid will have 1500 beams~\citep{L1_reqs}
      and MeerKAT 400 beams (B. Stappers
      priv. comm.).}}\label{fig:gain}
  \end{center}
\end{figure*}

Figure~\ref{fig:gain} shows the gain of SKA1 for sub-arrays consisting
of the inner 1~km, 20~km (and the full array for reference) for an
`average' line of sight through the Galaxy; for comparison several
other pulsar-search relevant telescopes are also
shown. Figure~\ref{fig:gain} shows the equal-bandwidth non-accelerated
pulsar survey speed figure of merit, defined as
$(A_{\mathrm{eff}}/T_{\mathrm{sys}})^2\Omega$. A number of caveats
must be made in interpreting these figures of merit. A survey
performed with sub-arrays of radii $\sim130$~m and $\sim500$~m could
be read as equal --- the former configuration involves less longer
pointings, the latter more shorter pointings. However this is not true
as the observing time per pointing is related directly to the
computing needed to process the data.
More importantly, this problem is compounded for acceleration searches
for two reasons: firstly one cannot compensate for less gain by
observing for longer as the number of compute operations required to
perform an acceleration search \textit{to find the same pulsar} scales
as $T_{\mathrm{obs}}^3$.
Secondly acceleration searches are ineffectual when $T_{\mathrm{obs}}$
is more than $\sim 10\%$ of the orbital periods under
consideration. As $T_{\mathrm{obs}}$ is limited in this way SKA1-Mid
pulsar searches will not use sub-arrays at the peak of the survey
speed figure of merit, but somewhat larger sub-arrays.

\subsection{Pulsar Timing Capabilities}
Pulsar timing capabilities are provided by the pulsar timing backend
(PST), a sub-element of CSP. For pulsars with DM in the range $0$ to
$3000$~$\mathrm{pc}\;\mathrm{cm}^{-3}$ and for periods as fast as
$0.4$~ms, it can perform observations with durations up to 5
hours. The beamformer produces up to 16 polarisation-calibrated
voltage beams for pulsar timing purposes (8 for SKA1-Mid Band 5). The
bandwidth of these beams is the entire 300-MHz band for SKA1-Low; for
SKA1-Mid it is the full band for Bands 1 and 2, and $2.5$~GHz for Band
5a/b. These data are coherently dedispersed in PST and folded
according to the known ephemerides. The process of `timing' to check
and update pulsar ephemerides, and react accordingly, is performed in
SDP. One can also perform pulsar searches using the timing backend
using the ``dynamic spectrum mode'', essentially a mode where PST
doesn't fold the data, and reduces the time and/or frequency
resolution to keep the data rate within limits, and with coherent
dedispersion optional. This mode is appropriate for long targetted
search pointings, e.g. Sgr A*, globular clusters, supernova remnants
etc. Noteworthy, but on the negative side, is that Band 3 (S-band) is
not covered in the initial construction budget --- an upgrade focused
on pulsar timing would no doubt prioritise this~\citep{css16,sc17}. In
terms of timing precision, there is uncertainty in the common delay
centre of each SKA sub-array in the amount of $2$~ns ($1\sigma$), with
the SKA timescale linked to UTC with an uncertainty of $5$~ns
($1\sigma$) for SKA1-Mid; for SKA1-Low it is $10$~ns
($1\sigma$). These effects combine in quadrature to give $5.4$~ns and
$10.2$~ns for SKA1-Mid and -Low respectively. TOA uncertainty
decreases with increasing S/N until this intrumental limit is reached,
if the limit due to intrinsic pulse jitter (which is pulsar specific)
is not reached first.

\section*{Acknowledgements}
\scriptsize{EK would like to thank everyone involved with the meeting
  for their fantastic efforts in delivering a truly amazing week. EK
  looks forward to defending the title of ``Pulsar Pub Quiz Champion''
  in 10 years time, at a venue also hopefully without wifi.}


\begin{thebibliography}{}  

\scriptsize{

\bibitem[Bailes \etal\ (2017)]{utmost} {Bailes, M. \etal\ } 2017,
  \textit{PASA}, 34, e045.

\bibitem[Barr \etal\ (2013)]{htrun} {Barr, E. D. \etal\ } 2013,
  \textit{MNRAS}, 435, 2234.

\bibitem[Braun \etal\ (2015a)]{SKA_Book} {Braun, R., Bourke, T., Green,
  J. A., Keane, E. F, \& Wagg, J.} 2015, \textit{PoS}, Proceedings of
  Advancing Astrophysics with the Square Kilometre Array (AASKA14),
  174.

\bibitem[Braun \etal\ (2015b)]{L0_reqs} {Braun, R., Bourke, T., Green,
  J. A., Keane, E. F, \& Wagg, J.} 2015, SKA1 Level 0 Science
  Requirements, SKA-TEL-SKO-0000007 revision 2, 2015-10-28.

\bibitem[Braun \etal\ (2017)]{performance} {Braun, R., Bonaldi, A.,
  Bourke, T., Keane, E. F, \& Wagg, J.} 2017, Anticipated SKA1 Science
  Performance, SKA-TEL-SKO-0000818 revision 1, 2017-10-17.

\bibitem[Caiazzo \etal\ (2017)]{L1_reqs} {Caiazzo, M., \etal\ } 2017,
  SKA Phase 1 System Requirements Specification, SKA-TEL-SKO-0000008
  revision 11, 2017-07-31.

\bibitem[Cordes \etal\ (2006)]{palfa} {Cordes, J. M. \etal\ } 2006,
  \textit{ApJ}, 637, 446.

\bibitem[Cordes \etal\ (2016)]{css16} {Cordes, J. M., Shannon, R. M.,
  Stinebring, D. R.} 2016, \textit{ApJ}, 817, 16.

\bibitem[van Haarlem \etal\ (2013)]{vanhaarlem} {van Haarlem,
  M. P. \etal\ } 2013, \textit{A\&A}, 556, A2.



\bibitem[Keane \etal\ (2018)]{superb} {Keane, E. F. \etal\ } 2018,
  \textit{MNRAS}, 473, 116.

\bibitem[Keith \etal\ (2010)]{htrus} {Keith, M. J.} 2010,
  \textit{MNRAS}, 409, 619.


\bibitem[Li \& Pan (2016)]{li_pan} {Li, D., Pan, Z.} 2016,
  \textit{Radio Science}, 51, 1060.

\bibitem[Rajwade \& Lorimer (2017)]{rl17} {Rajwade, K. M., Lorimer,
  D. R.} 2017, \textit{MNRAS}, 465, 2286.

\bibitem[Shannon \& Cordes (2017)]{sc17} {Shannon, R. M., Cordes,
  J. M.} 2017, \textit{MNRAS}, 464, 2075.

\bibitem[Stappers \etal\ (2011)]{Stappers_2011} {Stappers,
  B. W. \etal\ } 2011, \textit{A\&A}, 530, A80.

\bibitem[Stovall \etal\ (2014)]{gbncc} {Stovall, K. \etal\ } 2014,
  \textit{ApJ}, 791, 67.



}
\end{thebibliography}
\end{document}